\documentclass[12pt]{article}

\begin{document}
\title{On the linearization of the generalized Ermakov systems}
\author{F.~Haas \\
Instituto de F\'{\i}sica, UFRGS\\
Caixa Postal 15051\\
91500--970 Porto Alegre, RS - Brazil\\
J.~Goedert\\
Centro de Ci\^encias Exatas e Tecnol\'ogicas, UNISINOS\\
Av. Unisinos, 950\\
93022--000 S\~ao Leopoldo, RS - Brazil}
\date{\strut}
\maketitle

\begin{abstract}
A linearization procedure is proposed for Ermakov systems with frequency depending on dynamic variables. The procedure applies to a wide class of generalized Ermakov systems which are linearizable in a manner similar to that applicable to usual Ermakov systems. The Kepler--Ermakov systems belong into this category but others, more generic, systems are also included.

\end{abstract}

\section{Introduction}

Ermakov systems \cite{Ermakov}--\cite{Maamache} have merited special attention in recent years. A trend in the latest developments on the subject, is to focus attention in some special features of subclasses of Ermakov systems. These subclasses are frequently more flexible and may be tailored to suite some particular application or special purpose. Among others, we quote applications such as the identification of the Hamiltonian character in special circumstances \cite{Cervero, Haas}; the determination of a second constant of motion in certain other particular cases \cite{Goedert0}; the identification of the structure of associated Lie group of point symmetries \cite{Goedert, Leach} and the extension of the Ermakov systems concept itself to higher dimensions \cite{Leach}--\cite{Schief}.

A central feature of Ermakov systems is their property of always possessing a first integral, commonly known as the Lewis--Ray--Reid invariant.  The Lewis--Ray--Reid invariant can be used to construct nonlinear superposition laws, another important feature of Ermakov systems concerning their general solutions \cite{Reid}.

Recently, Athorne shown that Ermakov systems in their usual form, where the frequency function depends only on the time variable, are linearizable \cite{Athorne}. In the linearization process, the Lewis--Ray--Reid invariant plays a central role, analogous to that played by the angular momentum in the linearization of the two-body problem of classical mechanics \cite{Goldstein}. The linearization process, besides being an interesting mathematical attribute in itself, has found some important applications. The linearized form of the Ermakov systems plays a central role, for example, in the resolution of a problem arising in the theory of two--layer, shallow water waves \cite{Athorne1}, in the singularity analysis of Ermakov systems \cite{Athorne2, Athorne3} and in the study of higher dimensional Ermakov systems \cite{Rogers, Schief}. In such applications, however, a restricting condition on the class of
Ermakov system is that its frequency function depends only on time, a fact that imposes limitation on the scope of the linearization process. In a recent study, Athorne \cite{Athorne4} has found a class of dynamical systems, the so called Kepler--Ermakov systems, that generalizes the usual Ermakov systems while preserving the property of being amenable to linearization. The natural question to ask at this point is, therefore, whether the Kepler--Ermakov systems are the only linearizable generalization of the Ermakov systems or if  there exist other perturbations of usual Ermakov systems that are also amenable to linearization.

The frequency function entering the Ermakov system may depend on the dynamic variables and their derivatives, without any restriction on the existence of the  Lewis--Ray--Reid invariant \cite{Reid}. The resulting generalized Ermakov systems, with frequency function depending on the dynamic variables and their derivatives, besides time, have appeared in the study of a generalized time--dependent sine--Gordon equation \cite{Saermark}. More recently, the Hamiltonian character \cite{Haas}, the existence of a second exact invariant \cite{Goedert0} and the Lie point symmetry group \cite{Goedert} of Ermakov systems have been analyzed for such generalized frequency functions.  In the present work we address the question raised at the end of the previous paragraph and study the linearizability of generalized Ermakov systems where the frequency function may depend on the dynamic variables and their first derivatives, besides time. Our motivation is to find a more general class of dynamical systems still amenable to linearization, on the same line as the usual Ermakov or Kepler--Ermakov systems. In fact, as shown in detail in section III, the Kepler--Ermakov dynamical systems are nothing but a special class of the linearizable generalized Ermakov system.  This result puts the linearizability property  of Kepler--Ermakov systems in a novel, more generic and sound basis. It must be stressed, however, that the class of generalized linearizable Ermakov systems is much ampler then the class of Kepler--Ermakov systems, as will be seen in section II.

The paper is organized as follows. In section II, a linearization procedure is proposed for generalized Ermakov systems. The class of frequency functions for which the linearization procedure applies is obtained in terms of a wide category of generalized Ermakov systems that can be linearized. In the continuation, it is shown that the general solution for the nonlinear equations can be recovered from the solution for the corresponding linearized equation. The linearization of usual Ermakov systems is recovered as a special case of the more general theory. In section III, Kepler--Ermakov systems are shown to belong to the class of generalized linearizable Ermakov systems. The linearization procedure for the Kepler--Ermakov system  is illustrating in the case of a particular non-central force problem. In section IV an example of a generalized linearizable Ermakov system is presented which is {\it not} of the Kepler--Ermakov type. The linearization of this system is also exhibited. Section V is dedicated to the conclusions.

\section{Linearization of generalized Ermakov systems}

An Ermakov system \cite{Ray, Reid} in 2--D, is a pair of coupled, nonlinear second order differential equations,
\begin{eqnarray}
\label{er1}
\ddot x + \omega^{2}\,x &=& \frac{1}{yx^2}f(y/x) \,,\\
\label{er2}
\ddot y + \omega^{2}\,y &=& \frac{1}{xy^2}g(x/y) \,,
\end{eqnarray}
where $f$ and $g$ are arbitrary functions of their indicated arguments  and $\omega$, the so-called frequency function, may depend on time, the dynamic variables and their derivatives \cite{Reid}. In other words, $\omega =\omega(t,x,y,\dot{x},\dot{y},\ddot{x},\ddot{y}, \dots)$.  Either for physical or else for simplifying reasons, we restrict considerations to the cases where the frequency function depends at most on the velocity components. In this work, the nomenclature ``usual Ermakov system'' is used when $\omega = \omega(t)$, and ``generalized Ermakov system'', or Ermakov system for short, is used when the frequency has a more general dependence.

Independently of the way on which $\omega$ depends on the dynamic variables, the system (\ref{er1}--\ref{er2}) always possess the Lewis--Ray--Reid invariant
\begin{equation}
\label{er3}
I = \frac{1}{2}(x\dot y - y\dot x)^2 + U(y/x) \,,
\end{equation}
where $U$ is defined by
\begin{equation}
\label{er4}
U(y/x) = \int^{y/x}f(\lambda)\,d\lambda + \int^{x/y}g(\lambda)\,d\lambda \,.
\end{equation}
The invariance of $I$ can be directly verified by checking that $dI/dt = 0$ along any trajectory of the Ermakov system.

For the specific purpose of this work, polar coordinates $x = r\cos\theta$ and $y = r\sin\theta$ are more
appropriate. In this system of coordinates the Ermakov system becomes
\begin{eqnarray}
\label{er51}
\ddot r - r\dot\theta^2 + \omega^{2}r &=& F(\theta)/r^3 \,,\\
\label{er52}
r\ddot\theta + 2\dot{r}\dot\theta &=& - (dV/d\theta)/r^3 \,,
\end{eqnarray}
where $V(\theta) = U(\tan\theta)$, $F$ being defined in terms of $f$ and $g$ by
\begin{equation}
\label{er10}
F(\theta) = \frac{f(\tan\theta) + g(\cos\theta)}{\sin\theta\cos\theta} \,.
\end{equation}
In polar coordinates, the Lewis--Ray--Reid invariant becomes
\begin{equation}
\label{er5}
I = \frac{1}{2}(r^{2}\dot\theta)^2 + V(\theta) \,.
\end{equation}

As already pointed out before \cite{Haas,Goedert}, the concept of a generalized Ermakov system has originated from the observation that $\omega$ may depend arbitrarily on the dynamic  variables and that, as a consequence, only two and not three arbitrary functions are necessary to specify the Ermakov systems.  Indeed, redefining
\begin{equation}
\label{er50}
\omega^2 \mapsto \omega^2 - F(\theta)/r^4
\end{equation}
is equivalent to absorbing $F$ into $\omega$, that is, to making $F \equiv 0$ in the pair of equations (\ref{er51}--\ref{er52}).  Thus, strictly speaking generalized Ermakov system contains only two and not three arbitrary functions. This redefinition, however, is not essential in the present context and the usual notation is adopted in order to easy the comparison between our  results and the results found in the literature.

The linearization procedure for usual Ermakov systems is described in ref. \cite{Athorne}. Our strategy for the linearization of generalized Ermakov systems follows the same spirit of the procedure proposed for usual Ermakov systems. However, the assumption of a more generic dependence in $\omega$ suggests a straightforward generalization of the technique which leads to interesting new results.  In this way, we are able to find classes of generalized Ermakov systems, specified by appropriate dependence of $\omega$ in the dynamic variables, that are linearizable. For this purpose we follow Athorne \cite{Athorne} and introduce the new dependent variable $\psi$ defined by
\begin{equation}
\label{er6}
\psi = \rho(t)/r \,,
\end{equation}
where $\rho(t)$ is an arbitrary, unspecified function of time.  The new independent variable will be chosen as the angle $\theta$.  In order to obtain the equation of motion in the new variables, it is necessary to express the time variation of $\theta$ in terms of the invariant. This is achieved by use of the Lewis--Ray--Reid invariant (\ref{er5}), from which we construct
\begin{equation}
\label{er7}
\dot\theta = h(\theta;I)/r^2 \,,
\end{equation}
where a function $h,$ parametrically dependent on the numerical value of $I,$ was defined by the relation
\begin{equation}
\label{er8}
h(\theta;I) = \sqrt{2}(I - V(\theta))^{1/2} \,.
\end{equation}
At this point it is interesting to compare eq. (\ref{er7}) with the corresponding equation for $\dot\theta$ in the two body problem. Here, the Lewis--Ray--Reid invariant plays the role of angular momentum whereas $V(\theta)$ is a specific feature of the non-central nature of the motion.

Now using the Ermakov system (\ref{er51}--\ref{er52}) and  equations (\ref{er6},\ref{er8}), we easily arrive at the transformed equation
\begin{equation}
\label{er9}
h^{2}(\theta;I)\frac{d^{2}\psi}{d\theta^2} + h(\theta;I)\frac{\partial h(\theta;I)}{\partial\theta}\frac{d\psi}{d\theta}
 + \left(h^{2}(\theta;I) + F(\theta)\right)\,\psi = \frac{\rho^{3}(\ddot\rho + \omega^{2}\rho)}{\psi^3} \,.
\end{equation}
In order to obtain a linear equation in transformed variables, the dependence of $\omega$ in the dynamic variables must be chosen such that the right hand side of eq. (\ref{er9}) becomes a linear function of $\psi$ and $d\psi/d\theta$. This condition yields the compatibility condition for linearization,
\begin{equation}
\label{er60}
\frac{\rho^{3}(\ddot\rho + \omega^{2}\rho)}{\psi^3} = a(\theta;I)\frac{d\psi}{d\theta} + b(\theta;I)\psi + c(\theta;I)
\end{equation}
where $a$, $b$ and $c$ are arbitrary functions of the indicated arguments. A dependence on the numerical value of the Lewis--Ray--Reid invariant $I$ was included for maximal generality. Notice how the extra dependence in $\omega$ allows for a more general solution. In the case of $\omega=\omega(t)$ treated by Athorne, only $\ddot\rho+\omega^2\rho=0$ with $a \equiv b \equiv c \equiv 0$ was consistent with linearity. In the present case, $\rho$ remains arbitrary and three new functions $a$, $b$ and $c$ are introduced, leading to more general expressions for $\omega$ which are compatible with linearization. The resulting frequencies are given by
\begin{equation}
\label{er11}
\omega^2 = - \frac{\ddot\rho}{\rho} + \left(a(\theta;I)\frac{d\psi}{d\theta} + b(\theta;I)\psi + c(\theta;I)\right)\frac{\psi^3}{\rho^4} \,.
\end{equation}
In principle, we could include higher order derivatives in the right hand side of equation (\ref{er9}), without restrictions to linearity.  This possibility, however, was neglected, mainly for simplifying reasons.

Now restricting the frequency function of the form (\ref{er11}), we find
the linearized Ermakov equation
\begin{eqnarray}
h^{2}(\theta;I)\frac{d^{2}\psi}{d\theta^2} &+& \left(h(\theta;I)\frac{\partial h}{\partial\theta}(\theta;I) - a(\theta;I)\right)\frac{d\psi}{d\theta} + \nonumber \\
\label{er12}
&+& \left(h^{2}(\theta;I) + F(\theta) - b(\theta;I)\right)\,\psi = c(\theta;I) \,.
\end{eqnarray}
Frequencies not of the form (\ref{er11}), necessarily correspond to Ermakov equations that are not linear when the transformation $(r,t) \rightarrow (\psi,\theta)$ is performed.

In order to express the frequencies of the linearizable systems in terms of the original polar coordinates, we can use
\begin{equation}
d\psi/d\theta = \dot\psi/\dot\theta = - (\rho\dot r - \dot\rho r)/r^{2}\dot\theta \,,
\end{equation}
so that
\begin{equation}
\label{er13}
\omega^2 = - \frac{\ddot\rho}{\rho} + \frac{(\rho\dot r - \dot\rho r)}{\rho r^3}A(\theta, r^{2}\dot\theta) + \frac{B(\theta, r^{2}\dot\theta)}{r^4} + \frac{C(\theta, r^{2}\dot\theta)}{\rho r^3} \,,
\end{equation}
where
\begin{eqnarray}
\label{er14}
A(\theta, r^{2}\dot\theta) &=& - a(\theta;I)/(r^{2}\dot\theta) \,,\\
\label{er15}
B(\theta, r^{2}\dot\theta) &=& b(\theta;I) \,,\\
\label{er16}
C(\theta, r^{2}\dot\theta) &=& c(\theta;I)
\end{eqnarray}
are new arbitrary functions. Notice that the dependence of $A$, $B$ and $C$ in equations (\ref{er14}--\ref{er16}) is of the correct type, since the invariant (\ref{er5}) depends only on $\theta$ and $r^{2}\dot\theta$.

The resulting linearizable generalized Ermakov system in polar coordinates has the form
\begin{eqnarray}
\ddot r &-& r\dot\theta^2 - \frac{\ddot\rho}{\rho} r  + \frac{(\rho\dot r - \dot\rho r)}{\rho r^2}A(\theta, r^{2}\dot\theta) + \nonumber \\
\label{er70}
&+& \frac{B(\theta, r^{2}\dot\theta)}{r^3} + \frac{C(\theta,
r^{2}\dot\theta)}{\rho r^2}  = \frac{F(\theta)}{r^3} \,,\\
\label{er71}
r\ddot\theta + 2\dot{r}\dot\theta &=& - \frac{dV/d\theta}{r^3} \,,
\end{eqnarray}
and involves six arbitrary functions, namely $\rho, A, B, C, F$ and $V$. Other classes of generalized Ermakov systems amenable to linearization can exist, besides those generated by a dependence of $\omega$ on higher order derivatives. We may convince ourselves of this by noticing that $(\psi,\theta)$ is only one particular choice of transformed variables. Other different choices can be used, possibly leading to other  classes of linearizable generalized Ermakov systems. The great generality of the system (\ref{er70}--\ref{er71}), however, is already sufficient for the purposes of this paper.

Some additional remarks are in order here. First, as already pointed out, the function $F$ can be dropped from the system by an adequate redefinition of $\omega$. This is manifest in equation (\ref{er70}) where we see, by inspection, that $B$ can account for $F$. The second and perhaps more interesting remark is that the linearizable generalized Ermakov system (\ref{er70}--\ref{er71}) can be expressed in autonomous form, by means of a quasi--invariance \cite{Athorne, Burgan} transformation
\begin{equation}
\label{er80}
\bar{r} = r/\rho \quad , \quad \bar\theta = \theta \quad , \quad \bar{t} = \int^{t}d\lambda/\rho^{2}(\lambda) \quad .
\end{equation}
In these variables, we have the autonomous representation
\begin{eqnarray}
\bar{r}'' - \bar{r}\bar{\theta}'^2 + \frac{1}{\bar{r}^2}\left(A\bar{r}' + B/\bar{r} + C\right) &=& \frac{F(\bar\theta)}{\bar{r}^3} \,,\\
\bar{r}\bar{\theta}'' + 2\bar{r}'\bar{\theta}' = - \frac{dV(\bar{\theta})/d\bar{\theta}}{\bar{r}^3} \,,
\end{eqnarray}
where the prime stands for derivative with respect to $\bar{t}$ and $A$, $B$ and $C$ are functions of $\bar{r}^{2}\bar{\theta}'$ and $\bar\theta$. The time dependence in (\ref{er70}--\ref{er71}), is thus, in a sense, spurious. Consequently, there remains, in fact, only four fundamental arbitrary functions in the linearizable generalized Ermakov system (\ref{er70}--\ref{er71}), that is,  $F$ may be incorporated in $C$ and $\rho$ may be eliminated by the quasi--invariance transformation.  It must be stressed, however, that the quasi--invariance transformation is not an essential step in the linearization procedure.

We may now recover the solution for the nonlinear dynamics from the solution for the linearized Ermakov system. Let
\begin{equation}
\label{er79}
\psi = \psi(\theta; I,c_{1},c_{2}) = c_{1}\psi_{1}(\theta) + c_{2}\psi_{2}(\theta) + \psi_{p}(\theta)
\end{equation}
be the general solution for the linearized equation (\ref{er12}), where $c_1$ and $c_2$ are arbitrary constants, $\psi_1$ and $\psi_2$ are two linearly independent solutions for the homogeneous part of (\ref{er12}), and $\psi_p$ is any particular solution for (\ref{er12}). The solution for the nonlinear system follows from the definition (\ref{er6}) of the linearizing variable $\psi$ and relation (\ref{er7}), from which we find
\begin{equation}
\label{er85}
\dot\theta = h(\theta; I)\psi^{2}(\theta; I,c_{1},c_{2})/\rho^{2}(t) \,.
\end{equation}
This is a separable first order, ordinary differential equation, equivalent to the quadrature
\begin{equation}
\label{er250}
\int^{\theta}\frac{d\lambda}{h(\lambda;I)\psi^{2}(\lambda; I,c_{1},c_{2})} - \int^{t}\frac{d\lambda}{\rho^{2}(\lambda)} = J \,,
\end{equation}
where $J$ is the fourth integration constant of the equations of motion. The quadrature (\ref{er250}) locally yields the equivalent relation,
\begin{equation}
\label{er84}
\theta = \theta(t; I,J,c_{1},c_{2}) \,,
\end{equation}
which is the general solution for the angular variable, involving four arbitrary integration constants, or implicitly
\begin{equation}
\label{er86}
t = t(\theta; I,J,c_{1},c_{2}) \,.
\end{equation}
To obtain the orbits, we use the definition of $\psi$, the solution (\ref{er79}) for the linearized system and the relation
(\ref{er86}),
\begin{equation}
\label{er87}
r = r(\theta; I,J,c_{1},c_{2}) = \frac{\rho\left(t(\theta; I,J,c_{1},c_{2})\right)}{\psi(\theta; I,c_{1},c_{2})} \,.
\end{equation}

The time evolution of the radial variable follows from the definition of $\psi$, the solution for the linearized Ermakov system and the relation (\ref{er84}), from which we obtain
\begin{equation}
\label{er88}
r = r(t; I,J,c_{1},c_{2}) = \frac{\rho(t)}{\tilde\psi(t; I,J,c_{1},c_{2})} \,,
\end{equation}
where $\tilde\psi(t; I,c_{1},c_{2}) = \psi\left(\theta(t; I,c_{1},c_{2}); I,c_{1},c_{2}\right)$. While equation (\ref{er87}) represents the orbits for the nonlinear generalized Ermakov system, equations (\ref{er84}) and (\ref{er88}) give the time evolution of the dynamic variables, in terms of the general solution for the linearized system.

A particular case in the  class of generalized linearizable Ermakov systems (\ref{er70}--\ref{er71}), are the usual Ermakov systems, with frequency depending only on time, $\omega \equiv \omega(t)$. Let $A \equiv B \equiv C \equiv 0$ in the definition (\ref{er13}) of generalized frequencies, which implies
\begin{equation}
\label{er17}
\ddot\rho + \omega^{2}(t)\rho = 0 \,.
\end{equation}
Equation (\ref{er17}) is the equation for a harmonic oscillator \cite{Lewis0, Lewis} with time--dependent frequency.  In this particular case of  usual Ermakov system, the linearization (\ref{er12}) reduces to the form
\begin{equation}
\label{er19}
h^{2}(\theta;I)\frac{d^{2}\psi}{d\theta^2} + h(\theta;I)\frac{\partial h(\theta;I)}{\partial\theta}\frac{d\psi}{d\theta} + \left(h^{2}(\theta;I) + F(\theta)\right)\,\psi = 0 \,,
\end{equation}
which is a homogeneous linear equation. This result is in full agreement with that of Athorne \cite{Athorne}, although in a slightly different form, due to our particular choice of linearizing variables. Illustrative examples of linearization of usual Ermakov systems can be found in \cite{Athorne}, \cite{Athorne1}--\cite{Athorne4}. In the following section, a non-trivial generalized Ermakov system that was not, until now, treated as a member of the class of Ermakov systems is examined and its linearization explicitly calculated.

\section{Kepler--Ermakov systems}

Kepler--Ermakov systems \cite{Athorne4} are given by
\begin{eqnarray}
\label{er90}
\ddot r - r\dot\theta^2 &=& F(\theta)/r^3 - G(\theta)/r^2\,,\\
\label{er91}
r\ddot\theta + 2\dot{r}\dot\theta &=& - (dV(\theta)/d\theta)/r^3 \,,
\end{eqnarray}
with $F$, $G$ and $V$ are arbitrary functions of the indicated arguments. Kepler--Ermakov systems were introduced as a linearizable perturbation of usual Ermakov systems. The term $G$ in equation (\ref{er90}) destroys the usual Ermakov character of the system. However, the system (\ref{er90}--\ref{er91}) belongs to the class of linearizable generalized Ermakov systems derived in section II and consequently, qualifies as a generalized Ermakov system that is linearized by the procedure outlined above. This result sheds new light on the cause why Kepler--Ermakov systems are linearizable.  In fact, choose
\begin{equation}
\label{er20}
A = B = 0 \quad , \quad C = G(\theta) \quad , \quad \rho = 1
\end{equation}
in the definition of linearizable generalized Ermakov system  (\ref{er70}--\ref{er71}). This choice puts the system in the form of a Kepler--Ermakov system and the corresponding linearization is given by
\begin{equation}
\label{er23}
h^{2}(\theta;I)\frac{d^{2}\psi}{d\theta^2} + h(\theta;I)\frac{\partial h(\theta;I)}{\partial\theta}\frac{d\psi}{d\theta} + \left(h^{2}(\theta;I) + F(\theta)\right)\,\psi = G(\theta) \,,
\end{equation}
which, unlike the linearization of usual Ermakov systems, is a non-homogeneous linear equation. The non-usual character of the frequency function associated to Kepler--Ermakov systems is manifest in the equation
\begin{equation}
\omega^2 = G(\theta)/r^3 \,,
\end{equation}
which, evidently, shows a dependence on the dynamic variables.

A typical example of linearization of a Kepler--Ermakov system can be found in  reference \cite{Athorne4}. Other relevant systems that can be cast in the Kepler--Ermakov form are some of the superintegrable systems treated by Winternitz {\it et al.} \cite{Winternitz} and by Ra\~nada \cite{Ranada}. To inspect one of these systems as a Kepler--Ermakov system and to show how the linearization process works in a specific case, let us consider the Hamiltonian
\begin{equation}
\label{er200}
H = \frac{1}{2}\left(p_{r}^2 + \frac{p_{\theta}^2}{r^2}\right) - \frac{\mu_0}{r} + \frac{1}{r^2}\left(\frac{g_1 + g_{2}\cos\theta}{\sin^{2}\theta} + g_3\right) \,,
\end{equation}
where $\mu_0, g_1, g_2$ and $g_3$ are positive constants. When $g_3 \equiv 0$, the corresponding Hamilton--Jacobi equation is separable in parabolic and polar coordinates, a fact that  results from the  underlying dynamic symmetry algebra \cite{Winternitz}. For $g_3 \neq 0$, however, the system is still completely integrable, as shown in \cite{Haas}. Straightforward computation of the canonical equations
\begin{equation}
\dot r = \partial{H}/\partial p_r \quad , \quad \dot\theta = \partial{H}/\partial p_\theta \quad , \quad \dot p_r = - \partial{H}/\partial r \quad , \quad \dot p_\theta = - \partial{H}/\partial p_\theta
\end{equation}
shows that the Hamiltonian (\ref{er200}) yield a Kepler--Ermakov
system, with
\begin{eqnarray}
V(\theta) &=& (g_1 + g_{2}\cos\theta)/\sin^{2}\theta \,,\\
F(\theta) &=& 2 (V(\theta) + g_3) \,,\\
G(\theta) &=& \mu_0
\end{eqnarray}
in the equations of motion (\ref{er90}--\ref{er91}). The corresponding linear equation (\ref{er23}) is easier to treat in terms of a new time parameter $T(\theta; I)$ defined by
\begin{equation}
\label{er201}
T(\theta; I,J) = \int^{\theta}\frac{d\lambda}{h(\lambda; I)} + J \,.
\end{equation}
where $J$ is an arbitrary constant. Interestingly, this new time parameter depends parametrically on the numerical value of the Lewis--Ray--Reid invariant $I$. In terms of the new independent variable, the linearization becomes
\begin{equation}
\label{er203}
\frac{d^{2}\psi}{dT^2} + (I + g_{3})\psi = \mu_0 \,,
\end{equation}
the equation for a harmonic oscillator with a time--independent driving, which is always exactly solvable. The explicit form of the solution will depend  on the parameters $I$ and $g_3$. To construct  a particular case, take $I > 0$ and $g_3 > 0$. This choice together with (\ref{er203}) yields,
\begin{equation}
\label{er204}
\psi = \psi(T; c_{1},c_{2}) = c_{1}\cos(I + g_{3})^{1/2}T + c_{2}\sin(I + g_{3})^{1/2}T \,,
\end{equation}
where $c_1$ and $c_2$ are arbitrary constants. To express $\psi$  in terms of the angle $\theta$, we use  formula (\ref{er201}) which, for $I > 0$, furnishes
\begin{equation}
T(\theta; I,J) = -
\frac{1}{\sqrt{2I}}
\sin^{-1}\left(\frac{2I\cos\theta + g_2}
{(g_{2}^2 + 4I(I - g_{1}))^{1/2}}\right) + J\,.
\end{equation}
After substituting this result in  (\ref{er204}),  we find $\psi = \psi(\theta; I,J,c_{1},c_{2})$. The orbits are then  given by
\begin{equation}
r = r(\theta; I,J,c_{1},c_{2}) = 1/\psi(\theta; I,c_{1},c_{2}) \,,
\end{equation}
which is obtained from the definition of the linearizing variable $\psi$ and the fact that $\rho \equiv 1$ in the case of  Kepler--Ermakov systems. The actual form of the orbit is intricate and will be omitted here.  In spite of  the orbits being analytically given in the present example, the time evolution of the dynamic variables cannot be given analytically because the quadrature (\ref{er250}) is not  expressed in terms of elementary functions. Similar reasoning apply for other values of the parameters $I$ and $g_3$.

\section{Linearization and reduction to free motion}

In this section, we reverse the arguments and search for classes of Ermakov systems whose linearized form is simply a free motion. Among those, we find examples of linearizable Ermakov systems that are {\it not} of the  Kepler--Ermakov type. In addition we find that a whole class of Ermakov systems, still depending on two arbitrary functions, are reducible, by the linearization process, to simple free motion. Let
\begin{equation}
\label{er260}
a \equiv h\,\partial h/\partial\theta \quad , \quad b \equiv h^2 + F \quad , \quad c \equiv 0
\end{equation}
in the linear equation (\ref{er12}), yielding free particle motion,
\begin{equation}
d^{2}\psi/d\theta^2 = 0 \,.
\end{equation}
In this case the general solution is
\begin{equation}
\psi = c_1 + c_ {2}\theta \,,
\end{equation}
where $c_1$ and $c_2$ are arbitrary constants. In this case the  quadrature (\ref{er250}) has  the form
\begin{equation}
\label{er300}
\int^{\theta}\frac{d\lambda}{h(\lambda;I)(c_1 + c_{2}\lambda)^2} - \int^{t}\frac{d\lambda}{\rho^{2}(\lambda)} = J \,.
\end{equation}
For any given function $h(\theta; I)$, we locally find, from equation (\ref{er300}),  that either $t = t(\theta; I,J,c_{1},c_{2})$ or $\theta = \theta(t; I,J,c_{1},c_{2})$. The corresponding orbits are given by
\begin{equation}
r = r(\theta; I,J,c_{1},c_{2}) = \frac{\rho\left(t(\theta; I,J,c_{1},c_{2})\right)}{c_1 + c_{2}\theta} \,;
\end{equation}
whereas the time evolution of the radial variable follows from
\begin{equation}
r = r(t; I,J,c_{1},c_{2}) = \frac{\rho(t)}{c_1 + c_{2}\theta(t; I,J,c_{1},c_{2})}  \,.
\end{equation}
This completes, in the quadrature sense, the integration of the equations of motion.

Performing the steps in reversed order, we find, from (\ref{er14}--\ref{er16}) and (\ref{er260}), the dynamical system in the original variables
\begin{equation}
A = (dV/d\theta)/(r^{2}\dot\theta) \quad , \quad B = (r^{2}\dot\theta)^2 + F(\theta) \quad , \quad C = 0 \,\,.
\end{equation}
This yields the frequency function
\begin{equation}
\omega^2 = - \frac{\ddot\rho}{\rho} + \frac{(\rho\dot r - \dot\rho r)dV/d\theta}{\rho r^{5}\dot\theta} + \frac{(r^{2}\dot\theta)^2 + F}{r^4}
\end{equation}
and the corresponding linearizable generalized Ermakov system
\begin{eqnarray}
\rho\ddot r - \ddot\rho r &+& \frac{(\rho\dot r - \dot\rho r)}
{r^{4}\dot\theta}\frac{dV(\theta)}{d\theta} = 0 \,,\\
r\ddot\theta + 2\dot{r}\dot\theta &=& - \frac{dV(\theta)/d\theta}{r^3} \,.
\end{eqnarray}
As a final step, we can express the dynamical equations in cartesian form (\ref{er1}--\ref{er2}), with
\begin{eqnarray}
\omega^2 = - \frac{\ddot\rho}{\rho}
+ \left(\frac{x\dot y - y \dot x}{x^2 + y^2}\right)^2 &+&
\left(\frac{(\rho\dot x - \dot\rho x)x
+ (\rho\dot y - \dot\rho y)y}{\rho{x^{2}y^{2}}(x\dot y - y \dot x)}\right) f(y/x)\,,\\
g(x/y) &=& - f(y/x).
\end{eqnarray}
These systems which are of the generalized Ermakov type certainly
have remarkable properties: despite  having two arbitrary
functions, $\rho$ and $f$,  their solution can be reduced to the
equation of a free particle plus the quadrature (\ref{er300}).

\section{Conclusion}
In this paper, an extensive class of  generalized Ermakov systems, characterized by several arbitrary functions was shown to possess equivalent linear form. The linearization process depends basically on the existence of the Lewis--Ray-Reid invariant, a fact that is always assured for Ermakov systems, both in its usual or generalized form. As a particular result of the theory, it was shown  that the linearization of Kepler--Ermakov systems is nothing but a special case of the more general linearization of some classes of generalized Ermakov systems. While usual Ermakov systems are always linearizable, the same does not apply to generalized Ermakov systems. In the present work, use was made of the linearization variables $(\psi,\theta)$, the choice that originates the system (\ref{er70}--\ref{er71}). Different choices of linearizing variables could eventually lead to other classes of linearizable generalized Ermakov systems. As a final comments we remind that linearization could be a useful criteria to qualify or classify Ermakov systems. For instance, using the linearization, Athorne has classified usual Ermakov systems with respect to the rational character of their solutions \cite{Athorne2, Athorne3}. The same strategy can be applied to linearizable generalized Ermakov systems. This could be adopted as an alternative to the symmetry criteria proposed by Leach \cite{Leach} . At least the class of  linearizable Ermakov systems seems more extensive - four arbitrary functions - than the class of generalized Ermakov system with Lie point symmetry which is specified in terms of only two arbitrary functions.



\end{document}